\documentclass[apsrev4-1,prl,twocolumn,times,amsmath,amssymb,superscriptaddress,floatfix]{revtex4-1}

\usepackage{times}
\usepackage{float}
\usepackage{amsmath}
\usepackage{amsthm}
\usepackage{amssymb}
\usepackage{amsbsy}
\usepackage{graphicx}
\usepackage[colorlinks,bookmarks=false,citecolor=blue,linkcolor=red,urlcolor=blue]{hyperref}
\usepackage{pstricks}
\usepackage{color}
\usepackage{cancel}
\usepackage{subfigure} 
\usepackage{enumerate}
\usepackage{mathrsfs} 
\usepackage{multirow} 
\usepackage{ulem}
\usepackage{wasysym}
\usepackage{sidecap}

\begin{document}

\title{Instabilities of a U(1) quantum spin liquid in
 disordered non-Kramers pyrochlores}

\author{Owen Benton}
%
\affiliation{RIKEN Center for Emergent Matter Science (CEMS), Wako, Saitama, 351-0198,
Japan}

\begin{abstract}
Quantum spin liquids  (QSLs) are exotic phases of matter exhibiting long-range
entanglement and supporting emergent gauge fields.
A vigorous search for experimental realizations of these states
has identified several materials with properties hinting at QSL physics.
A key issue in understanding these QSL candidates is often the
interplay of weak disorder of the crystal structure with the spin liquid state.
It has recently been pointed out that in at least one important class of candidate 
QSLs - pyrochlore magnets based on non-Kramers ions such as Pr$^{3+}$
or Tb$^{3+}$- structural disorder can actually promote a $U(1)$ QSL ground state.
Here we set this proposal on a quantitative footing by analyzing the stability 
 of the QSL state in the minimal
model for these systems: a random transverse field Ising model.
We consider two kinds of instability, which are relevant in different limits of the
phase diagram: condensation of spinons and confinement
of the $U(1)$ gauge fields.
Having obtained stability bounds on the QSL state we apply our results directly to the
disordered candidate QSL Pr$_2$Zr$_2$O$_7$.
We find that the available data for currently studied samples of 
Pr$_2$Zr$_2$O$_7$ is most consistent with it
a ground state outside the spin liquid regime,
in a paramagnetic phase
 with quadrupole moments near saturation
due to the influence of structural disorder.
\end{abstract}

\maketitle

Experimental realizations of Quantum Spin Liquid (QSL) 
states are the goal of a long-running 
research effort~\cite{lee08, balents10}.
Interest in QSLs stems from their ability to support
fractional excitations, emergent gauge fields and large-scale
quantum entanglement \cite{savary17-RMP80, zhou17}.
Several candidate QSLs are known and one key subset of these
is found amongst 
pyrochlore oxides R$_2$M$_2$O$_7$
\cite{gardner10, gingras14}.
The geometrical frustration of the pyrochlore lattice famously gives 
rise to spin ice- a classical spin liquid with magnetic monopole
excitations-  in Ho and Dy based pyrochlores 
\cite{harris97, castelnovo08, castelnovo12}.
The theoretical result that a spin ice imbued with
quantum fluctuations can host a $U(1)$ QSL
with emergent photons \cite{hermele04, banerjee08, savary12,
shannon12, benton12, hao14, kato15, mcclarty15, chen16} has fueled interest in
spin-ice-like systems
with stronger quantum effects
\cite{ross11, chang12, tokiwa15, thompson-arXiv,
fennell12, kermarrec15, takatsu16, hallas15, sibille15, zhou08, kimura13, 
petit16-NatPhys12, 
sibille16, anand16, sibille17-arXiv}.

A recurrent issue in these investigations is the role of quenched 
disorder \cite{ross12, taniguchi13, arpino17, mostaed17}.
Recently \cite{savary17-PRL118}, Savary and Balents have demonstrated that for 
pyrochlores based on non-Kramers ions weak structural
disorder can actually promote the QSL ground state.
This is because structural imperfections around the
 magnetic sites act as transverse fields on the low energy effective
$S=1/2$ degrees of freedom.
These transverse fields induce tunneling
between classical spin ice ground states, which stabilizes the $U(1)$ QSL.
Experiments confirm the presence
of these transverse fields in the candidate quantum spin ice Pr$_2$Zr$_2$O$_7$
\cite{wen17, martin17}
suggesting the possibility of a 
disorder-induced QSL ground state.

\begin{figure}[h!]
\centering
\includegraphics[width=\columnwidth]{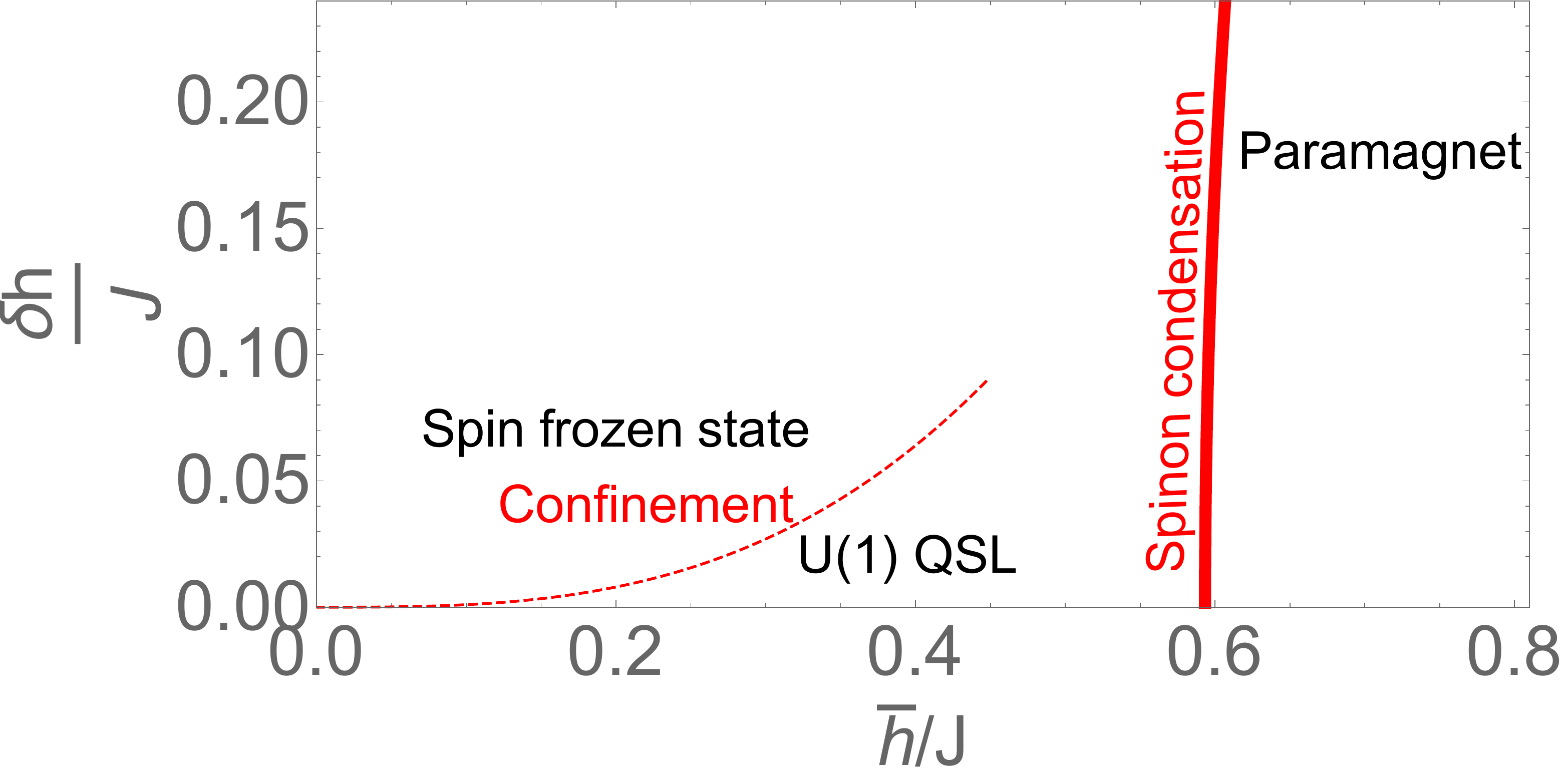}
\caption{Instabilites of the $U(1)$ QSL in the random transverse field Ising
model on the pyrochlore lattice [Eq. (\ref{eq:HRTFIM})].
There are two instabilities which appear
upon increasing the average strength $\bar{h}$ and 
width $\delta h$ of the distribution of transverse fields.
The condensation of spinons leads to a
trivial paramagnetic
phase, and the threshold for this instablity can be calculated
by calcularing the spinon energy perturbatively
[Eq. (\ref{eq:stability})].
The second instability is to confinement of the $U(1)$ gauge field,
leading to a glassy state with frozen magnetic moments.
The threshold for this instability occurs along a line $\delta h = \alpha \bar{h}^3$,
for small $\delta h/J,\bar{h}/J$.
The coefficient $\alpha$ depends on the details of the
distribution of transverse fields, the figure shows a sketch with $\alpha=1$. 
}
\label{fig:stability}
\end{figure}

This Letter addresses two questions.
Firstly, what is the extent of the QSL phase
in the minimal model for non-Kramers pyrochlores with
weak structural disorder?
Secondly, do currently studied samples of Pr$_2$Zr$_2$O$_7$ fall within this
QSL phase?

The first of these questions is answered by
considering two instabilities of the $U(1)$ QSL:
spinon condensation and confinement.
The threshhold for each can be calculated
in perturbation theory.
Fig. \ref{fig:stability} shows the 
results of this calculation.

Determining to which phase Pr$_2$Zr$_2$O$_7$ belongs
requires parameterizing a model for this material.
We do this by comparing
available thermodynamic data to
 Numerical Linked Cluster (NLC) \cite{rigol06, rigol07, tang13}
calculations.
Our model suggests current samples of
 Pr$_2$Zr$_2$O$_7$ fall within a
paramagnetic phase, with $4f$ quadrupole
moments nearly saturated by the effective transverse fields.
Exact Diagonalization (ED) calculations suggest this conclusion is
consistent with scattering experiments showing
an excitation continuum and broadened, spin-ice-like, correlations at
low energies \cite{kimura13, petit16-PRB94}.

{\it Stability regime of the QSL}- We consider 
a minimal model for non-Kramers pyrochlores
where the degeneracy of the ground crystal electric field (CEF) 
doublet
is lifted by local deviations from $D_{3d}$ site symmetry.
We assume that the gap to higher energy CEF states is large
such that the only relevant degrees of freedom are Pauli
matrices $\vec{\sigma}_i$ describing the two
states of the ground doublet.
Due to the symmetry of non-Kramers doublets \cite{onoda10, onoda11} on
the pyrochlore lattice, the magnetic moment on site $i$
points only along the local-axis $\hat{{\bf z}_i}$ joining the centers
of the pyrochlore tetrahedra sharing the site
\begin{eqnarray}
{\bf m}_i=\mu_{\sf eff} \sigma^z_i \hat{\bf z}_i
\label{eq:moment}
\end{eqnarray}
where $\mu_{\sf eff} $ is the effective moment size.
Nearest-neighbor Ising interactions
\begin{eqnarray}
\mathcal{H}_{\sf SI}=J\sum_{\langle ij \rangle} \sigma^z_i \sigma^z_j
\label{eq:HSI}
\end{eqnarray}
with $J>0$ favour spin-ice-like states in which the total value of
$\sigma^z_i$ vanishes on every tetrahedron in the lattice
\begin{eqnarray}
\sum_{i \in t} \sigma^z_i=0 \quad \forall \ \text{tetrahedra} \ t
\label{eq:icerule}
\end{eqnarray}

The transverse pseudospin operators $\sigma^{x, y}_i$ are
time-reversal invariant and cannot
contribute to the magnetic moment.
A finite value of these operators corresponds instead to a
finite quadrupole moment  \cite{onoda10, onoda11}.
The time-reversal invariance of $\sigma^{x, y}_i$ 
allows them to couple linearly to lattice imperfections 
which lift the local $D_{3d}$ symmetry \cite{savary17-PRL118}.
These imperfections thus act as a transverse field on $\sigma_i$
\begin{eqnarray}
\mathcal{H}_{\sf TF}= - \sum_i h_i \sigma^x_i
\label{eq:HTF}
\end{eqnarray}
where we have used local coordinate transformations on $
\sigma^{\alpha}_i$ such that the coupling is always to 
$\sigma^x_i$ \cite{savary17-PRL118}.
The transverse fields $h_i$ are distributed on the interval $[0, \infty]$
and  we  assume them to be 
uncorrelated in space 
$
\overline{h_i h_j}=\overline{h_i} \ \ \overline{h_j}
$.
We use $\overline{x}$ to denote the average of $x$ over
disorder realizations, reserving $\langle x \rangle$ for quantum statistical averages
at fixed disorder realization.

The minimal model for non-Kramers pyrochlores with
weak structural disorder is thus a random transverse
field Ising model  \cite{savary17-PRL118}
\begin{eqnarray}
\mathcal{H}_{\sf RTFIM}= \mathcal{H}_{\sf SI}+\mathcal{H}_{\sf TF}.
\label{eq:HRTFIM}
\end{eqnarray}

For weak, uniform, $h_i$ the ground state is a $U(1)$ QSL,
while for $h_i \gg J$ a  trivial paramagnetic  state is expected 
\cite{savary17-PRL118}.
The transition between these phases occurs via condensation 
of the gapped spinon excitations of the $U(1)$ QSL \cite{savary17-PRL118, roechner16}.
Below, we use perturbation theory to estimate the threshhold for this transition.
We then consider an alternative, confinement, instability of the QSL
which leads to a frozen moment state.

In the limit $h_i=0$ a spinon corresponds to a tetrahedron $t$
where Eq. (\ref{eq:icerule}) is violated,
with
$
\sum_{i \in t} \sigma^z_i=\pm 2
$
and a gap
$
\Delta_0=2 J
$.
To calculate the gap 
in the presence of disordered transverse fields we 
consider a state containing $M$ spinons. We take $1 \ll M \ll N_t$,
where $N_t$ is the number of tetrahedra in the lattice, such that the spinon density
is low and spinon interactions may be neglected.
Using second order perturbation theory we obtain an effective Hamiltonian
acting amongst these $M$
spinon states
\begin{eqnarray}
&&\mathcal{H}_{\sf eff}^{(M)}=E_0^{\sf cl}+ M \Delta_0+ 
\mathcal{H}_{1}^{(M)}+\mathcal{H}_{2}^{(M)}
\label{eq:HeffM}
\\
&&
\mathcal{H}_{1}^{(M)}=
\mathcal{P}_M
\mathcal{H}_{\sf TF}
\mathcal{P}_M 
\label{eq:H1M}
\\
&& 
\mathcal{H}_{2}^{(M)}=
-
\mathcal{P}_M
\mathcal{H}_{\sf TF}
\frac{\mathcal{Q}_M}{\mathcal{H}_{\sf SI}-(E_0^{\sf cl}+ M \Delta_0)}
\mathcal{H}_{\sf TF}
\mathcal{P}_M \quad
\label{eq:H2M}
\end{eqnarray}
where $E_0^{\sf cl}=-NJ$, $\mathcal{P}_M$
projects onto the manifold of states with $M$ spinons and $\mathcal{Q}_M$
projects onto its orthogonal complement.

To find the lowest energy state for
$M$ spinons we use the fact that for  $1 \ll M \ll N_t$ the 
column sum of $\mathcal{H}_{\sf eff}^{(M)}$ is approximately constant
\cite{kourtis16, wan16}.
To see this, consider first $\mathcal{H}_1^{(M)}$, which allows
each spinon to hop to three of its four neighboring tetrahedra.
The column sum of $\mathcal{H}_1^{(M)}$ is
\begin{eqnarray}
\sum_{\alpha} \left(\mathcal{H}_1^{(M)} \right)_{\alpha \beta}=-\sum_{i \in {\text{flippable}}} h_i
\end{eqnarray} 
For sparse spinons $M \ll N$ there are $3M$ flippable spins and
\begin{eqnarray}
\sum_{\alpha} \left(\mathcal{H}_1^{(M)} \right)_{\alpha \beta}=-3 M \frac{1}{3M}\sum_{i \in {\text{flippable}}} h_i=-3M \overline{h}
\label{eq:hop1}
\end{eqnarray} 
where we have used the fact that $M\gg1$ and the
assumption that $h_i$ are uncorrelated on different sites.

The second order part of the effective Hamiltonian  $\mathcal{H}_2^{(M)}$,
contains a diagonal contribution which is constant for $1 \ll M \ll N$
\begin{eqnarray}
\left(\mathcal{H}_2^{(M)} \right)_{\alpha \alpha}=-\frac{N_t}{2J} \overline{h^2} + \frac{7M}{8J}\overline{h^2} 
\label{eq:diag2}
\end{eqnarray}
coming from virtual processes which flip the same spin twice.
The off-diagonal part of $\left(\mathcal{H}_2^{(M)} \right)_{\alpha \beta}$ 
enables spinons to hop to 6 out of their 12 second-nearest-neighbor
tetrahedra by flipping two spins $k, l$ with matrix element $-\frac{h_k h_l}{4 J}$.
Since  each $M$ spinon configuration can
tunnel to the same number of other $M$ spinon configurations we find that 
the column sum of the second order Hamiltonian is also approximately constant:
\begin{eqnarray}
\sum_{\alpha} \left(\mathcal{H}_2^{(M)} \right)_{\alpha \beta}=-\frac{N_t}{2J} \overline{h^2} + \frac{7M}{8J}\overline{h^2} -\frac{3M}{2J}\overline{h}^2
\label{eq:hop2}
\end{eqnarray} 

Since  $\mathcal{H}_{\sf eff}^{(M)}$ [Eq. (\ref{eq:HeffM})] has approximately constant column
sum, and negative off-diagonal matrix elements its 
ground state must be an equal weight, Rokhsar-Kivelson-like, superposition of all configurations 
containing $M$ spinons \cite{kourtis16, wan16, rokhsar88}
\begin{eqnarray}
|\phi_M\rangle=\frac{1}{\sqrt{\mathcal{N}_M}} 
\sum_{|\alpha\rangle\in  \{|M\rangle \}} |\alpha\rangle
\end{eqnarray}
where $\mathcal{N}_M$ is the number of such configurations.

Using this  wave function  and Eq. (\ref{eq:HeffM})
to calculate the energy gives
\begin{eqnarray}
&& E(M)= 
E_0^{\sf cl} -\frac{N_t \overline{h^2} }{2J} 
+M\left(2J -3 \overline{h} +\frac{7 \delta h^2-5\overline{h}^2}{8J}
\right) \ \ \ \quad
\label{eq:energy-M}
\end{eqnarray}
where 
$
\delta h=\sqrt{\overline{h^2}-\overline{h}^2}
$.

When the coefficient of $M$ in Eq. (\ref{eq:energy-M})
becomes negative it becomes favorable
for spinons to  proliferate and condense.
Stability of the QSL against spinon condensation thus requires
\begin{eqnarray}
2J -3 \overline{h} +\frac{7}{8J} \delta h^2
-\frac{5\overline{h}^2}{8 J}>0
\label{eq:stability}
\end{eqnarray}
as plotted in Fig. \ref{fig:stability}.
Beyond this line 
the system gives way to a trivial paramagnetic
ground state.

\begin{figure}
\centering
\subfigure[\ Heat capacity ]{
\includegraphics[width=\columnwidth]{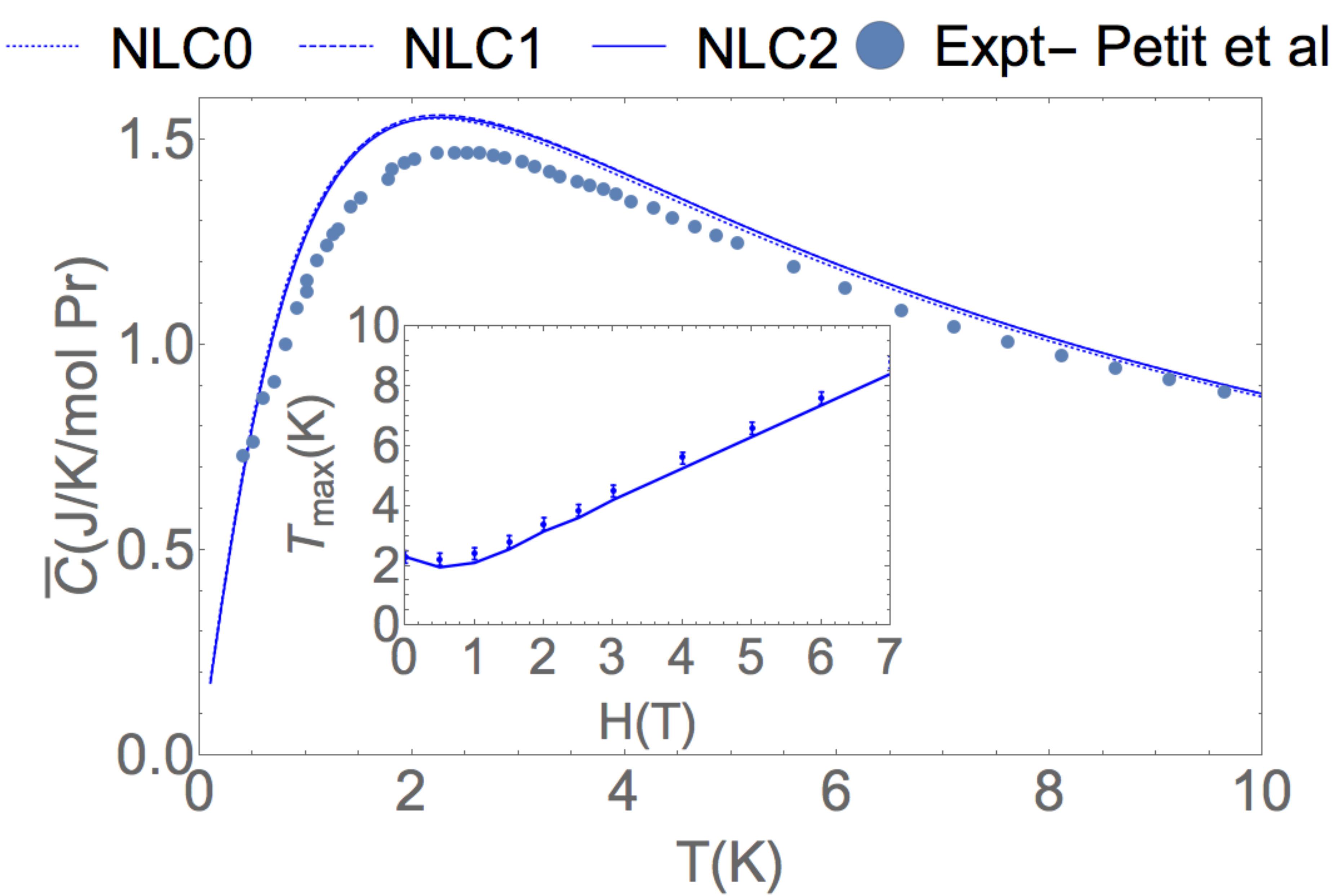}
\label{fig:cv}} \\
\subfigure[\ Inverse susceptibility ]{
\includegraphics[width=\columnwidth]{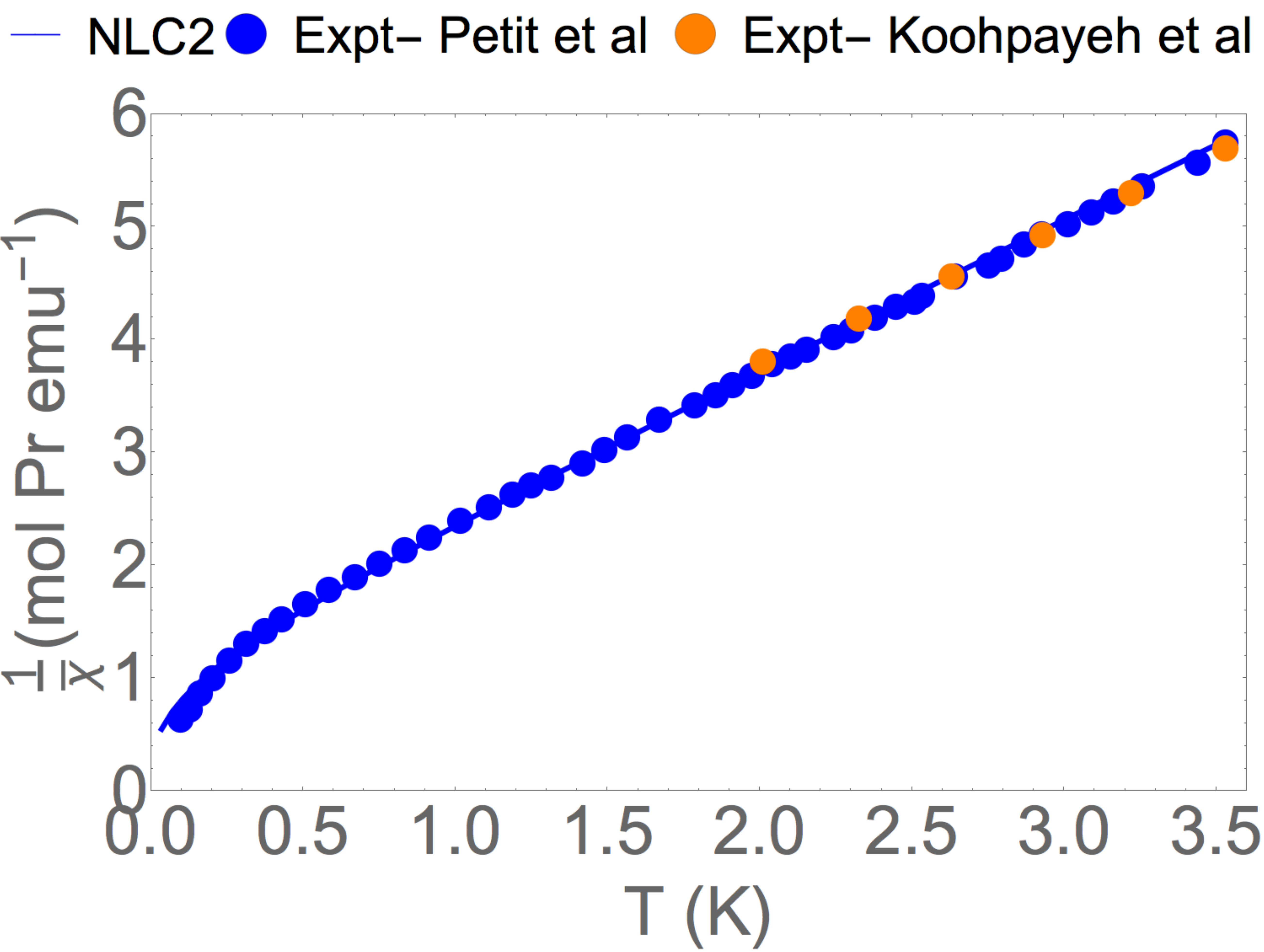}
\label{fig:chi}} 
\caption{Establishing a model for Pr$_2$Zr$_2$O$_7$.
NLC calculations for the random 
transverse field Ising model [Eq. (\ref{eq:HRTFIM})]
using $J=0.02$ meV,  effective moment
$\mu_{\sf eff}=2.45 \mu_B$
 and a Lorentzian distribution of transverse
fields with width $\Gamma=0.2$ meV give a good
description of the heat capacity [(a)] and inverse
susceptibility  [(b)].
Experimental results for the heat capacity are extracted \cite{footnote1} from
Ref. \cite{petit16-PRB94} and results for the susceptibility from 
Refs. \cite{petit16-PRB94, koohpayeh14}.
The inset of (a) shows the development of the temperature of 
the specific heat maximum under a magnetic field applied along
the [110] direction, compared between NLC calculations and
data from Ref. \cite{petit16-PRB94}.
The heat capacity of isostructural La$_2$Zr$_2$O$_7$ \cite{matsuhira09} was subtracted from the 
data from \cite{petit16-PRB94} to remove the phonon contribution.
Calculations are averaged over $10^6$ disorder realizations.
}
\label{fig:ModellingPZO}
\end{figure}

We can compare the result of Eq. (\ref{eq:stability}) to  
the phase boundary of the uniform transverse field
model ($\delta h=0$) found in Ref. \cite{roechner16}.
Inserting $\delta h=0, \overline{h}=h$ into Eq. (\ref{eq:stability}) we find the
critical value for $h$ is $h_c\approx0.593 J$, close to the
$h_c\approx0.602 J$ obtained from high field expansion in \cite{roechner16}.
The agreement with the results of  \cite{roechner16}
for the uniform case suggests that the second order calculation is sufficient,
at least when $\delta h$ is small.

In deriving Eq. (\ref{eq:stability})
we have considered $M$ spinon states with $1 \ll M \ll N$,
to justify  inserting averaged matrix elements in 
Eqs. (\ref{eq:hop1})-(\ref{eq:hop2}).
One may wonder what happens for states with small spinon number $M\sim1$.
In this case a lower energy may be obtained by restricting spinons to small 
subregions with untypically large values of $h_i$.
The condensation of spinons within these subregions can thus occur before the
bulk instability predicted by Eq. (\ref{eq:stability}), leading to a
Griffiths phase in which most of the system remains in the QSL
state but there are rare paramagnetic regions \cite{savary17-PRL118}.
The Griffiths phase is difficult to address analytically but could be 
be identified in simulation via spinon zero modes 
appearing at the boundaries of the paramagnetic
regions \cite{savary17-PRL118}.

While Eq. (\ref{eq:stability}) establishes a bulk stability 
bound against spinon condensation, there is another relevant
instability for the QSL.
This second instability corresponds to the condensation of the ``magnetic"
monopole charge, leading to a confinement transition \cite{chen16}.
Unlike the spinons, the magnetic monopole is an excitation within the
ice manifold, so this instability should be addressed using degenerate 
perturbation theory within the classical ground states.
A perturbative treatment of $\mathcal{H}_{\sf TF}$ within the
ground state manifold of $\mathcal{H}_{\sf SI}$ leads to a term
at fourth order which makes the 
Ising exchange interactions bond dependent
\begin{eqnarray}
J \to J+\delta J_{ij}, \ \delta J_{ij}=\frac{h_i^2 h_j^2}{48 J^3}
\end{eqnarray}
(see Supplemental Material for details).
The ground state of the  bond dependent Ising Hamiltonian
will be some frozen state of $\sigma_i^z$ which depends on the
disorder realization.
Such a ground state corresponds to a confined phase of 
the $U(1)$ gauge theory.

In the limit of uniform transverse fields ($\delta h=0$) this term becomes a constant
 and the leading non-trivial term is then a sixth order ring
exchange $g\sim h^6$ which stablizies the $U(1)$ QSL. 
At finite $\delta h$, the transition between the QSL and confined
phases must occur when $g\sim \delta J$, which gives a phase boundary
\begin{eqnarray}
\delta h = \alpha \bar{h}^3
\end{eqnarray}
with $\alpha$ depending on the details of the transverse field distribution.
Determination of $\alpha$ for a given type of distribution requires a numerical study beyond
the scope of this work.

{\it Modeling Pr$_2$Zr$_2$O$_7$}- We now seek to 
establish a model for the candidate 
QSL Pr$_2$Zr$_2$O$_7$,  and determine its
location on the phase diagram.
In Ref. \cite{wen17} the distribution of $h_i$
 arising in a sample of Pr$_2$Zr$_2$O$_7$ was characterized 
by analyzing inelastic neutron scattering results
in an applied magnetic field. The result was a  Lorentzian distribution
\begin{eqnarray}
p(h)=\frac{2\Gamma}{\pi} \frac{1}{\Gamma^2+h^2}, \quad h \in [0, \infty]
\label{eq:lorentzian}
\end{eqnarray}
with $\Gamma=0.27$meV.

Inspired by this we have compared  thermodynamic
data from other samples of Pr$_2$Zr$_2$O$_7$ \cite{petit16-PRB94, bonville16, koohpayeh14}
to NLC calculations using the Hamiltonian in Eq. (\ref{eq:HRTFIM}), with a Lorentzian
distribution of transverse fields (see Supplemental Material for details).
The NLC expansion is a means of estimating quantities in the thermodynamic limit from 
a series of diagonalizations of small clusters \cite{rigol06, rigol07, tang13}, 
which has been used successfully for other pyrochlores
\cite{singh12, applegate12, hayre13, jaubert15}.
Disorder averages can be taken term by term in the expansion \cite{tang15-PRB91, tang15}.
Calculations are done using zeroth (NLC0), first (NLC1) and
second (NLC2) order expansions, incorporating clusters of 1 site, 
1 tetrahedron and two tetrahedra respectively.
The interaction strength $J$, distribution width $\Gamma$ and effective moment
$\mu_{\sf eff}$ are treated as adjustable
parameters.
We have focussed on obtaining agreement with 
thermodynamic data from Refs. \cite{petit16-PRB94, bonville16},
but the quantitatively similar heat capacity curves obtained 
elsewhere \cite{kimura13, koohpayeh14} suggest that the model we obtain
should be approximately valid for other currently studied samples.

Agreement with heat capacity and susceptibility data is 
obtained in the parameter range $J=(0.020 \pm 0.005) $meV
$\Gamma=(0.20\pm0.01)$meV, 
$\mu_{\sf eff}=(2.45 \pm 0.05) \mu_B$, as shown in Figure \ref{fig:ModellingPZO}(a)-(b).
Our fits capture the antiferromagnetic Curie-Weiss behavior in spite of having a spin-ice like $J>0$.
They also capture the broad maximum in the specific heat and its evolution as a function of
applied field [inset of Fig. \ref{fig:cv}].

\begin{figure}
\includegraphics[width=\columnwidth]{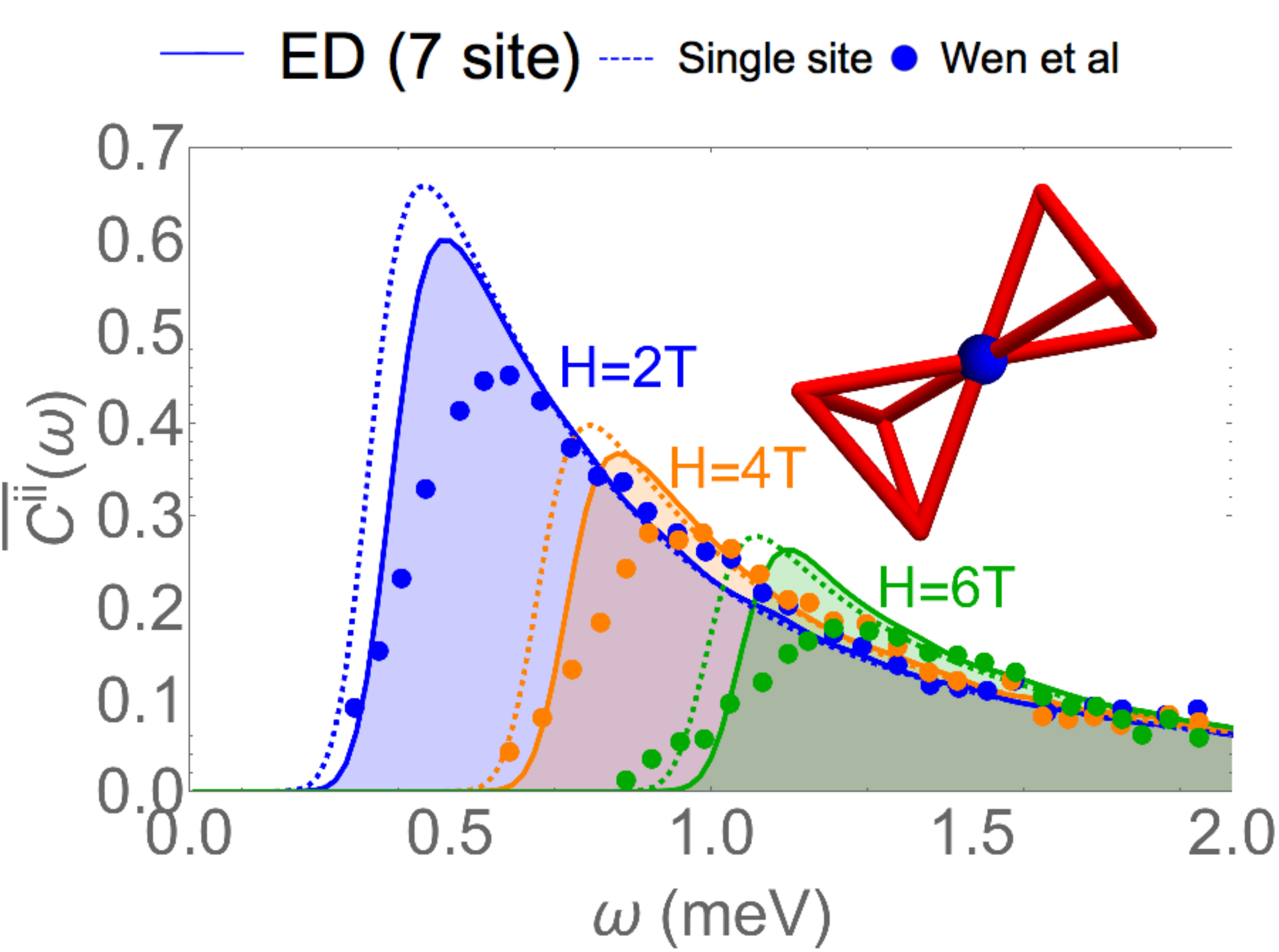}
\caption{%
On-site correlation function $\overline{C}_{ii}(\omega)$ calculated for
three values of external magnetic field along the [100] direction.
Calculations are made using ED on the 7-site cluster 
shown in the inset, using parameters $J=0.02$meV
and $\Gamma=0.2$meV.
$C_{ii}(\omega)$  is calculated for the central spin of the cluster.
The dashed lines show a single site calculation (i.e. neglecting interactions $J$)
and the calculations are compared with ${\bf q}$ integrated data from Ref. \cite{wen17},
which are multiplied by an overall scale factor.
The calculation has been
 averaged over $2\times10^5$ disorder realizations
and
 convoluted with a Gaussian of Full Width at Half Maximum (FWHM)
=0.11 meV to mimic experimental resolution.
}
\label{fig:scattering100}
\end{figure}

Despite obtaining a  narrower distribution of transverse fields than quoted
in \cite{wen17}, our fitted model gives a reasonable description of the
neutron scattering data from that study.
This is shown in Fig. \ref{fig:scattering100}, where we compare the disorder averaged on-site 
correlation function
\begin{eqnarray}
\overline{C}_{ii}(\omega)=\overline{\sum_{|\alpha \rangle}| \langle0|\sigma^z_i |\alpha\rangle|^2\delta(\omega-E_{\alpha})}
\end{eqnarray}
for the central spin of a 2-tetrahedron cluster with ${\bf q}$-integrated scattering data from
Ref. \cite{wen17}.
The  model  overestimates the scattering close to the Zeeman energy at each
value of field, but agrees closely with the high energy scattering, and agrees qualitatively with
the form of the lower energy scattering.
Differences between the model and experimental data may be attributable to
interactions not included in the simple model Eq. (\ref{eq:HRTFIM}),
spatial correlations in the transverse field distribution \cite{wen17} 
and variation between samples.

What does this model suggest about Pr$_2$Zr$_2$O$_7$?
A difficulty with the Lorentzian distribution [Eq. (\ref{eq:lorentzian})]
is that its moments $\overline{h}, \overline{h^2}$  are not well
defined.
This inhibits direct application of the stability criterion (\ref{eq:stability}).
We can circumvent this issue by applying a finite cut-off $h_{max}$
to the distribution in Eq. (\ref{eq:lorentzian}) and observing the
trajectory of $\overline{h}, \delta h$ as the cut-off is increased, while
keeping $J=0.02$meV and $\Gamma=0.2$meV.
Upon the increasing the cut-off from $h_{max}=0$, the model
crosses into the paramagnetic region of Fig. \ref{fig:stability}
for cut-offs as low as $h_{max, c}=0.025$meV.
Since the distribution of transverse fields in Pr$_2$Zr$_2$O$_7$
certainly extends far beyond this point, we should expect  Pr$_2$Zr$_2$O$_7$
to fall deep within the paramagnetic phase of the model.
This agrees with both NLC and 16-site ED calculations
which predict a nearly saturated ground state expectation value of
$\overline{\langle \sigma_x \rangle}\approx0.98$ with  $J=0.02$meV
and $\Gamma=0.20$meV.

\begin{figure}
\centering
\includegraphics[width=0.7\columnwidth]{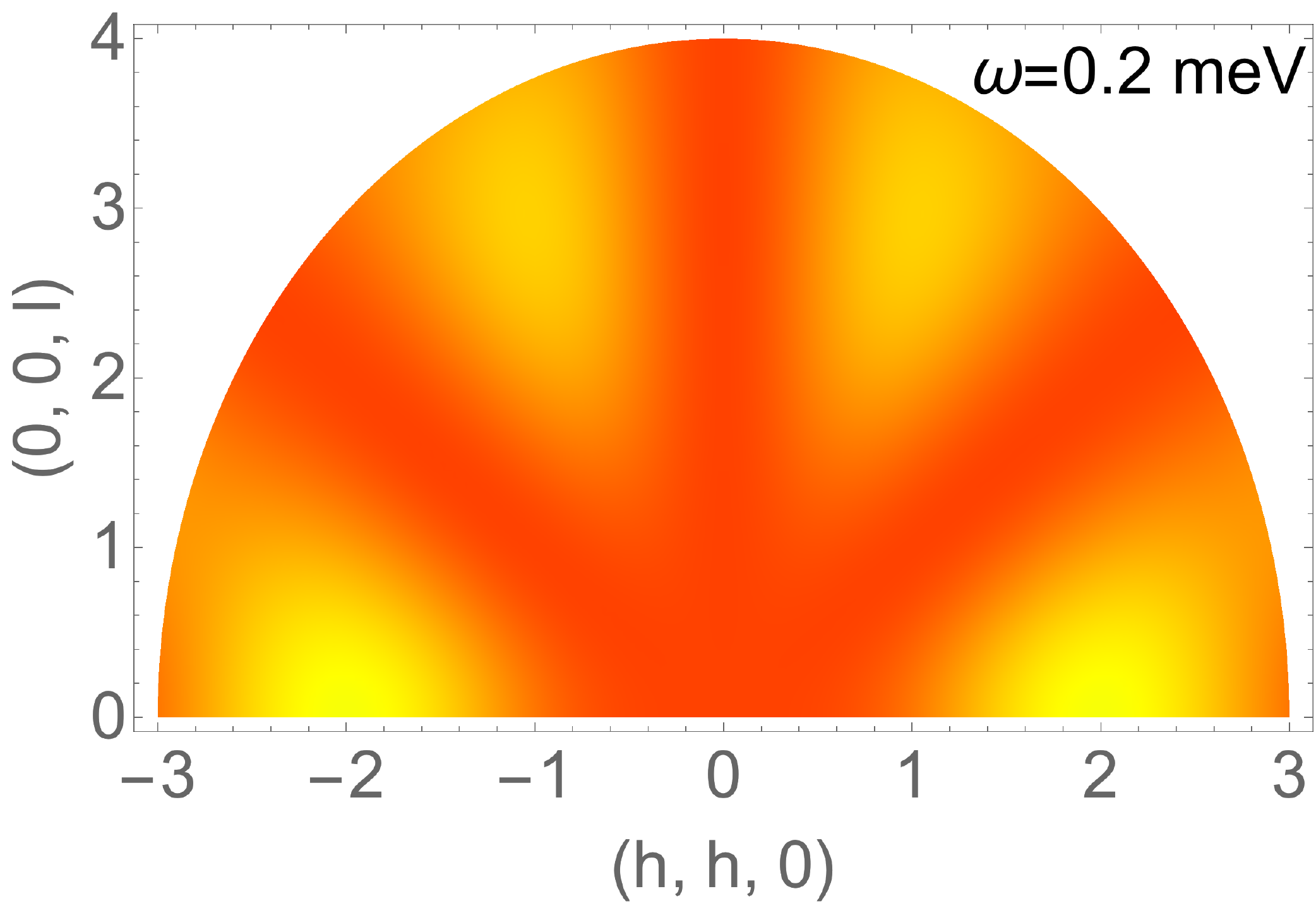}
\includegraphics[width=0.085\columnwidth]{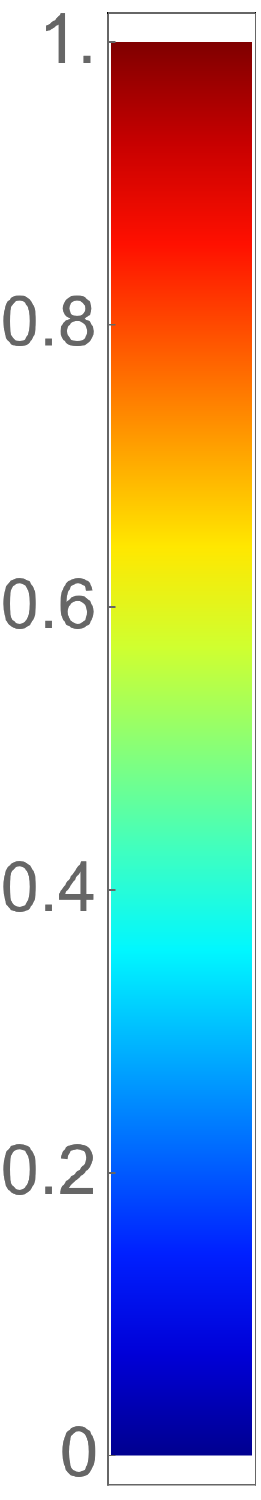}
\includegraphics[width=0.085\columnwidth]{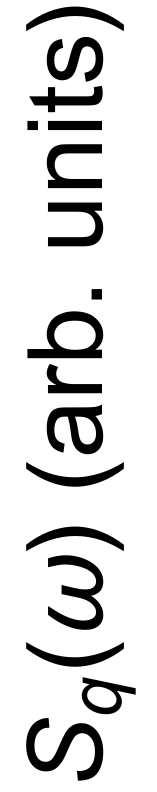}
\caption{Calculation of the inelastic structure factor at $\omega=0.2$meV
for Pr$_2$Zr$_2$O$_7$ [Eq. (\ref{eq:Sqw1}) ].
The calculation uses real space correlators
$ \overline{C}_{ij}(\omega) $ calculated from Exact Diagonalization of 
Eq. (\ref{eq:HRTFIM}) on a 16 site cubic cluster,
with $J=0.02$meV
and a Lorentzian distribution of transverse fields [Eq. (\ref{eq:lorentzian})] with
$\Gamma=0.2$meV.
Calculations are averaged over 300 realisations of disorder and
convoluted with a Gaussian of FWHM
=0.11 meV to mimic finite experimental resolution.
This calculation reproduces the broadened
remnants of spin-ice like correlations observed at finite energy
in Pr$_2$Zr$_2$O$_7$ \cite{kimura13, petit16-PRB94}. 
}
\label{fig:Scattering}
\end{figure}

A natural question at this point is whether this conclusion can
be reconciled with inelastic neutron scattering
experiments \cite{kimura13, petit16-PRB94}.
To address this we have calculated disorder averaged 
real space
correlation functions, $\overline{C}_{ij}(\omega)$,  up to third nearest 
neighbour in 16-site ED and combined them into a calculation of the 
dynamical structure factor
for neutron scattering
\begin{eqnarray}
S_{\mathbf{q}}(\omega)=
\sum_{i, j}
{\frac{e^{i \mathbf{q} \cdot  ({\bf r}_i-{\bf r}_j)} 
\left( \hat{\bf z}_i \cdot \hat{\bf z}_j 
- (\hat{\bf z}_i \cdot \hat{\bf q}) (\hat{\bf z}_j \cdot \hat{\bf q}) 
\right)
\overline{C}_{ij}(\omega) 
}{N}} \qquad
\label{eq:Sqw1}
\end{eqnarray}
Using our model parameters to calculate this at finite energy $\omega=0.2$meV, we obtain a similar
pattern to that observed in Refs. \cite{kimura13, petit16-PRB94}, namely broadened
remnants of spin-ice like correlations.
This suggests that neutron scattering observations on Pr$_2$Zr$_2$O$_7$ can 
be reconciled with the paramagnetic state predicted here. 

{\it Conclusions}-
We have investigated the instabilities of the $U(1)$ QSL
against spinon condensation and confinement
in a model describing non-Kramers pyrochlore magnets
with weak disorder.
We have parameterized this model for currently studied samples of
the  Pr$_2$Zr$_2$O$_7$ and found that they most likely
fall within the paramagnetic regime, a result consistent
with available scattering data.
An interesting direction for future research is to seek 
control of the transverse field distribution by varying experimental 
parameters in the synthesis procedure.
If one can tune through the spinon condensation
threshhold in this way, then not only the $U(1)$ QSL
phase but also the topological quantum phase transition
connecting the QSL to the paramagnetic phase become accessible.
The determination of the phase boundary and the method of estimating model
parameters from thermodynamics used in this work
can aid  in the fine tuning of samples through
the phase transition.

Other Pr pyrochlores
such as Pr$_2$Hf$_2$O$_7$ \cite{sibille16, anand16} and Pr$_2$Sn$_2$O$_7$ \cite{zhou08}
are of great interest, particularly given
recent experimental results indicating the possible existence
of emergent photons in Pr$_2$Hf$_2$O$_7$ \cite{sibille17-arXiv}.
Further work is needed to determine
whether these materials realize the
$U(1)$ QSL.
%

{\it Acknowledgments} 
The author acknowledges useful discussions with Bella Lake, Kate Ross,
Alexandros Samartzis, Nic Shannon and Jiajia Wen.


\section{Supplemental Material: Instabilities of a $U(1)$ quantum spin liquid in
disordered non-Kramers pyrochlores}

\section{Details of Numerical Linked Cluster Calculations}

Here we give some details of the Numerical Linked Cluster (NLC) 
calculations presented in the main text.
A pedagogical introduction to NLC expansions is given in \cite{tang13}.

In Numerical Linked Cluster expansions an extensive quantity $\mathcal{O}$
divided by the number of sites $N$, is calculated as a sum over
contributions from all clusters $c$ that can be embedded in the lattice
\begin{eqnarray}
\frac{1}{N}\langle \mathcal{O} \rangle = \sum_{c } M(c) W(c).
\end{eqnarray}

$M(c)$ is the multiplicity of the cluster per site- i.e. how many times that cluster
can be embedded in a lattice of $N$ sites, divided by $N$.
$W(c)$ is the cluster weight defined as
\begin{eqnarray}
W(c)=\langle \mathcal{O} \rangle_c-\sum_{s \subset c} W(s)
\label{eq:NLCweight}
\end{eqnarray}
where $\langle \mathcal{O} \rangle_c$ is the expectation value of
$\mathcal{O}$ on the cluster $c$, which is calculated from exact diagonalization.
The sum in the second term is a sum of the weights of all the subclusters
of $c$.

\begin{figure}[h!]
\centering
\subfigure[\ Cluster $c_1$: One site]{
\includegraphics[width=0.1\textwidth]{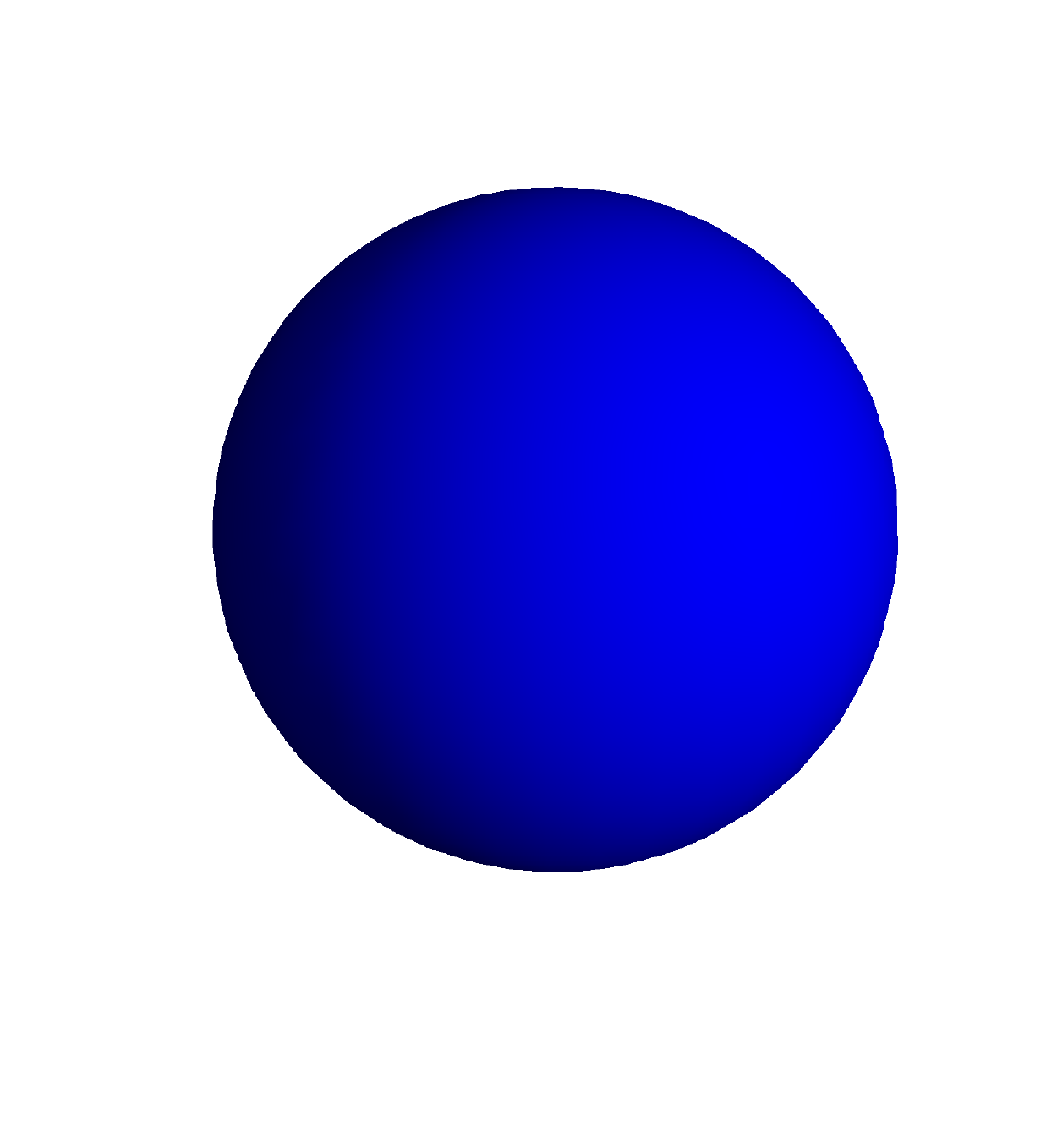}
\label{fig:nlc0}} 
\subfigure[\ Cluster $c_4$: One tetrahedron]{
\includegraphics[width=0.14\textwidth]{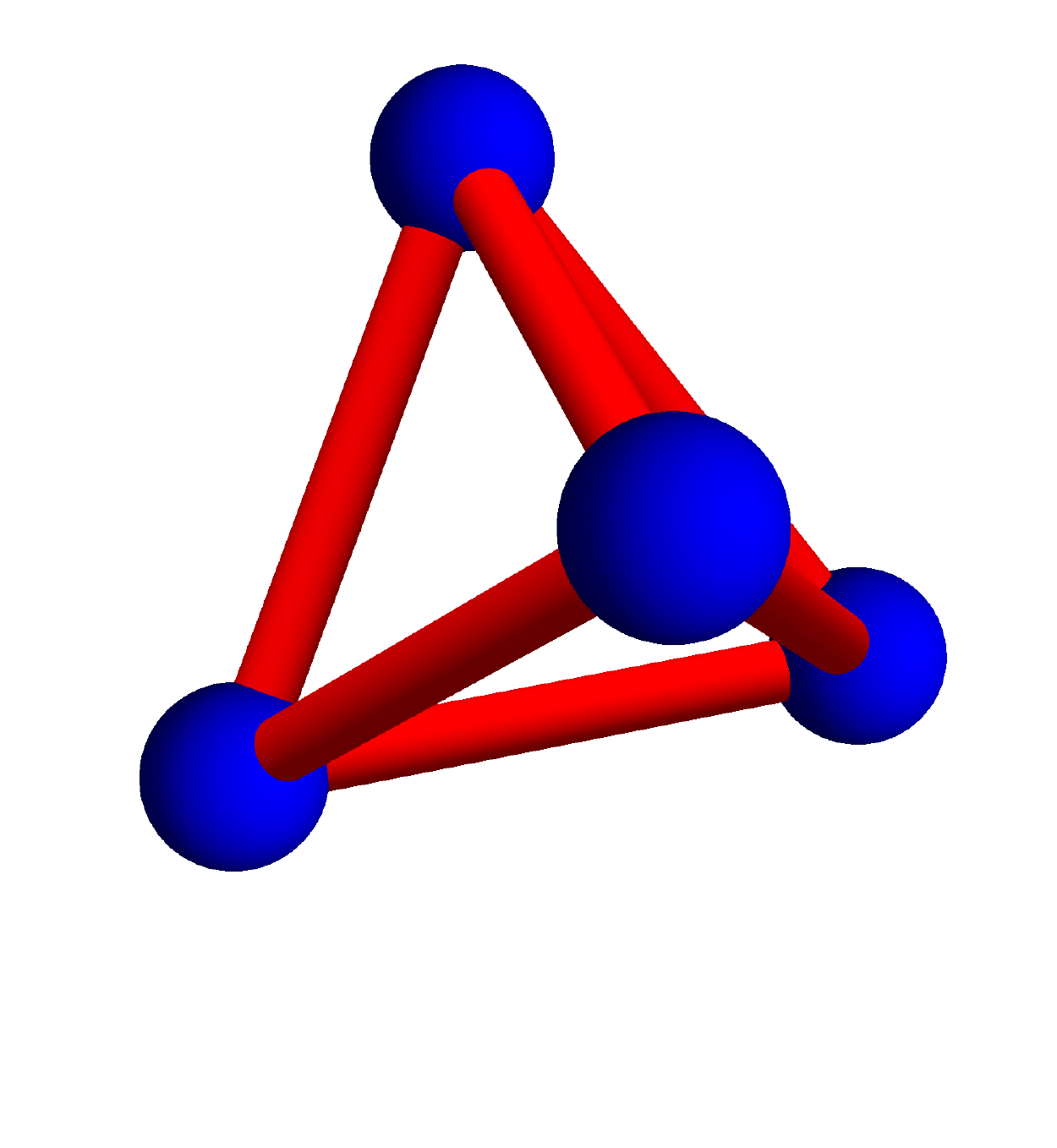}
\label{fig:nlc1}}  
\subfigure[\ Cluster $c_7$: Two tetrahedra]{
\includegraphics[width=0.2\textwidth]{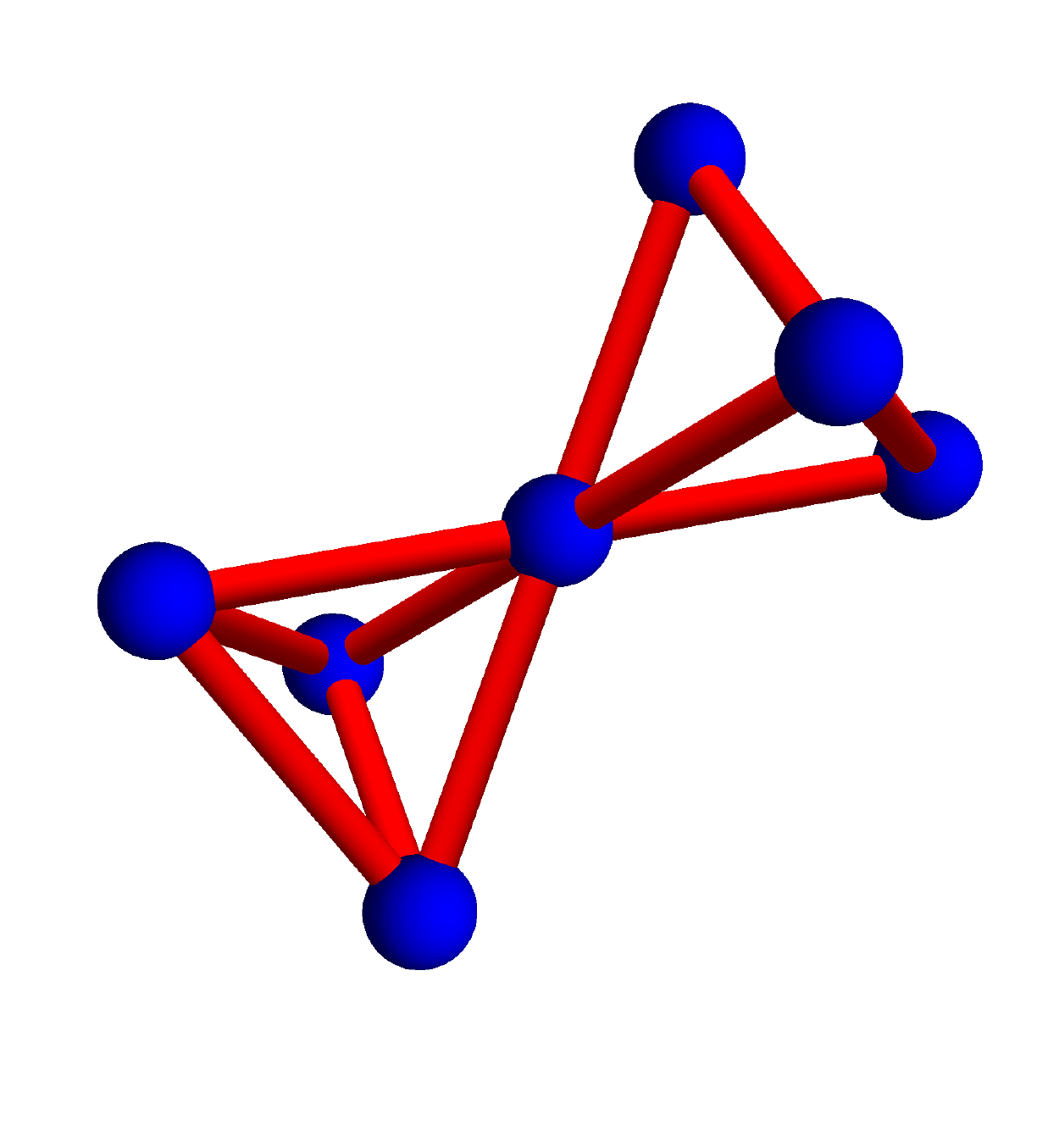}
\label{fig:nlc2}}  
\caption{Series of clusters used in NLC calculations in
the main text.}
\label{fig:clusters}
\end{figure}

In our calculations we have used the series of clusters shown in 
Fig. \ref{fig:clusters}, calculating the series up to second order.
We have used clusters of 1, 4 and 7 sites and we denote them as
$c_1, c_4, c_7$.
The multiplicities of these clusters per site  are
\begin{eqnarray}
M(c_1)=1, \quad M(c_4)=\frac{1}{2} \quad M(c_7)=1
\end{eqnarray}
and the weights for calculation of quantity $\mathcal{O}$ per site
are
\begin{eqnarray}
&&W(c_1)=\langle\mathcal{O} \rangle_{c_1} \nonumber \\
&&W(c_4)=\langle\mathcal{O} \rangle_{c_4}-4\langle\mathcal{O} \rangle_{c_1}
\nonumber \\
&&W(c_7)=\langle\mathcal{O} \rangle_{c_7}
-2(\langle\mathcal{O} \rangle_{c_4}-4\langle\mathcal{O} \rangle_{c_1})
-7 \langle\mathcal{O} \rangle_{c_1}.
\end{eqnarray}

In the presence of disorder, disorder averaged quantities
can be calculated by taking the disorder average term by term \cite{tang15}, i.e.
\begin{eqnarray}
&&\frac{1}{N}\overline{\langle \mathcal{O} \rangle} = \sum_{c } M(c) \overline{W(c)} \\
&&\overline{W(c)}=\overline{\langle \mathcal{O} \rangle_c}-\sum_{s \subset c} \overline{W(s)}
\end{eqnarray}

For the calculations in the presence of an external [110]
magnetic field [inset of Fig. 2(a)
of main text], the reduction in point group symmetry due to the applied field means that the 
clusters $c_1$ and $c_7$ now have two inequivalent types which must be treated separately \cite{hayre13}.

\section{Optimization of model parameters for $\text{Pr}_2\text{Zr}_2\text{O}_7$}

Here we describe the procedure used to optimize the model parameters
$J, \Gamma, \mu_{\sf eff}$ to describe the thermodynamics of $\text{Pr}_2\text{Zr}_2\text{O}_7$
[Fig. 2 of main text].

\begin{figure}
\centering
\qquad \ \ \includegraphics[width=0.34\textwidth]{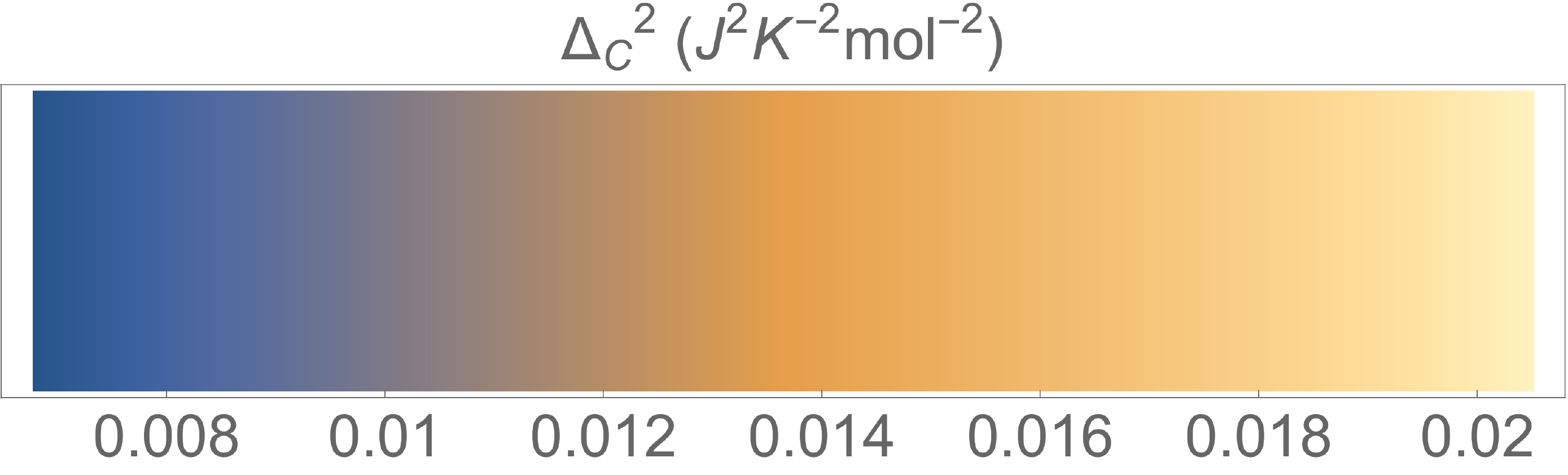}\\
\includegraphics[width=0.4\textwidth]{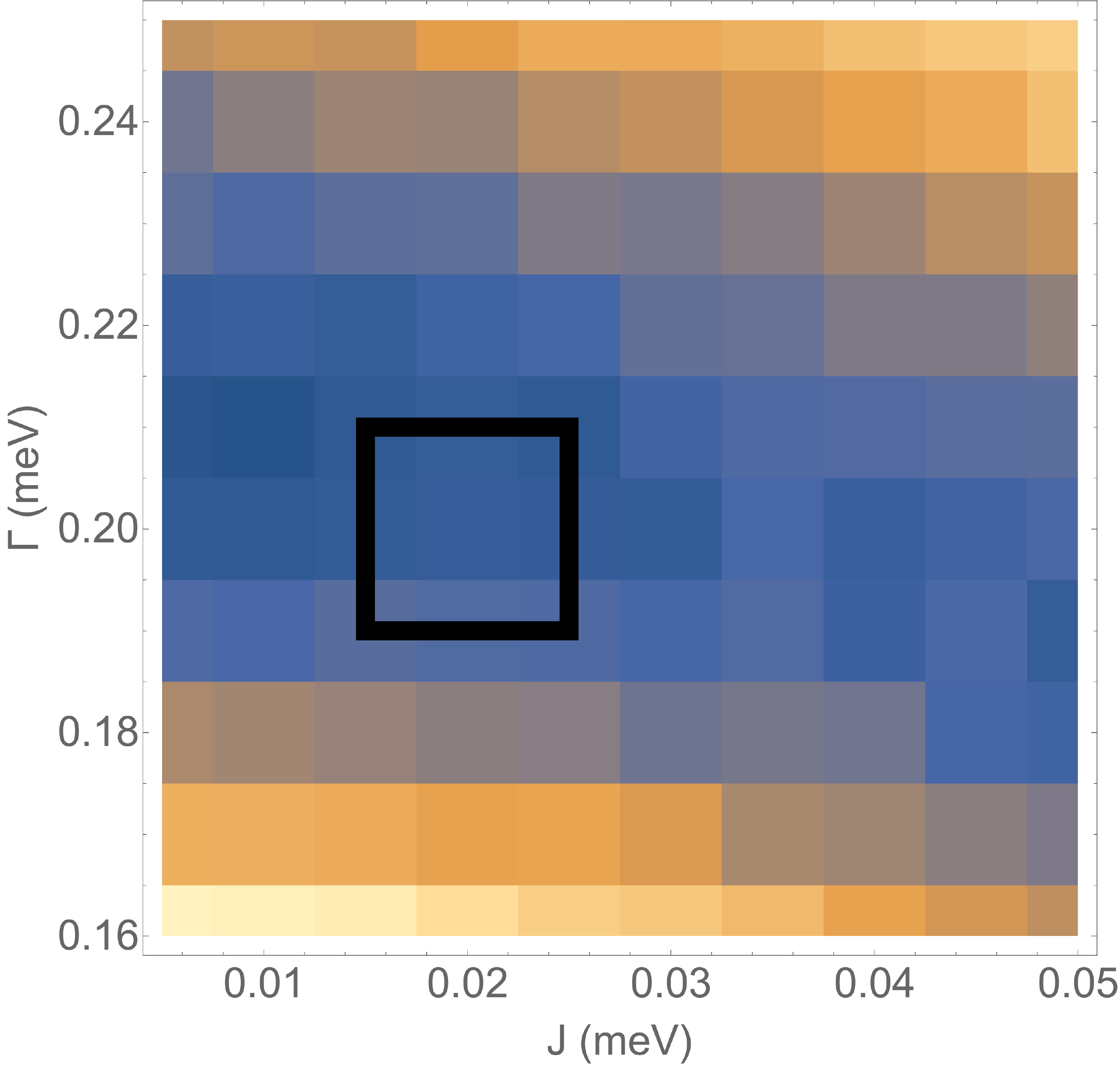}
\caption{Total squared error of the fit to the zero-field heat capacity data of Ref. 
[\onlinecite{petit16-PRB94}], as a function of the width parameter $\Gamma$
and the Ising exchange $J$. 
Calculations were made for each value of $\Gamma, J$ using second-order NLC
expansion and averaging over $10^6$ realizations of disorder.
The best fits are obtained with $\Gamma \in [0.19, 0.22] $ meV,  $J<0.035$meV. 
In this limit the fit quality is only weakly dependent on $J$ which must be fixed using 
the susceptibility data [see Fig. \ref{fig:chi_quality}].
The black square indicates the region where t
where both$(\Delta_{C})^2$ and  $(\Delta_{\chi^{-1}})^2$
are minimized, up to the accuracy of our calculations.
This region constitutes our estimates of the model parameters [Eq. (\ref{eq:estim})].
 }
\label{fig:C_quality}
\end{figure}

To begin with we consider the zero-field heat capacity data in the temperature range 
$T \in [0.4, 10]$K,
which we have extracted from Ref. [\onlinecite{petit16-PRB94}].
For each value of the temperature we have subtracted 
the lattice specific heat based on the measurements
for non-magnetic La$_2$Zr$_2$O$_7$ in \cite{matsuhira09}, to obtain the
experimental magnetic heat capacity $C^{\sf mag, exp} (T)$.

We then calculate $\bar{C} (J, \Gamma, T)$  in second order NLC 
for a series of values
of $J$ at intervals of $0.005$meV and $\Gamma$ at intervals of 
$0.01$ meV respectvely.
The calculation is made using disorder averaging over $10^6$
realizations of disorder.
Note that the zero field heat capacity is independent of the effective 
moment $\mu_{\sf eff}$.

For each value of $J$, $\Gamma$ we calculate the total squared error
\begin{eqnarray}
\Delta_C^2=\frac{1}{N_p}\sum_i (C^{\sf mag, exp}_i (T_i)- \bar{C} (J, \Gamma, T_i)   )^2
\label{eq:TSE_C}
\end{eqnarray}
where the index $i$ runs over experimental data points
$C^{\sf mag, exp}_i (T_i)$ and $N_p=47$ is the number of data points used.

The results of this calculation are shown in Fig. \ref{fig:C_quality}.
The best fits are obtained with distribution widths $\Gamma \in [0.19, 0.22]$meV,
and with $J<0.035$meV.
This is a parameter regime where the heat capacity is dominated by the
distribution of transverse fields, so the quality of fit is only very weakly dependent
on $J$.

To fix $J$ and $\mu_{\sf eff}$  we turn to the inverse susceptiility data.
Similarly to the treatment of the heat capacity we extract the 
experimental susceptibility $\frac{1}{\chi^{\sf exp} (T)}$ from
Ref.  \cite{petit16-PRB94}.
We then calculate $\bar{\chi}^{-1} (J, \Gamma, \mu_{\sf eff}, T)$  in second order NLC 
for the same parameter sets $J, \Gamma$ used to calculate the heat capacity in Fig.
\ref{fig:C_quality}, averaging over $10^6$
realizations of disorder..
Initially we set $\mu_{\sf eff}=2.45 \mu_B$ as found in Ref.  \cite{petit16-PRB94}.

The total squared error for the inverse susceptibility is
\begin{eqnarray}
\Delta_{\chi^{-1}}^2=\frac{1}{N_p}\sum_i 
\left(  \frac{1}{\chi^{\sf exp}_i} (T_i)- \bar{\chi}^{-1} (J, \Gamma, \mu_{\sf eff}, T_i)  \right )^2
\label{eq:TSE_C}
\end{eqnarray}
where $i$ runs over experimental data points
$\frac{1}{\chi^{\sf exp}_i (T_i)}$ and $N_p=51$ is the number of data points used.

\begin{figure}
\centering
\qquad \ \ \includegraphics[width=0.34\textwidth]{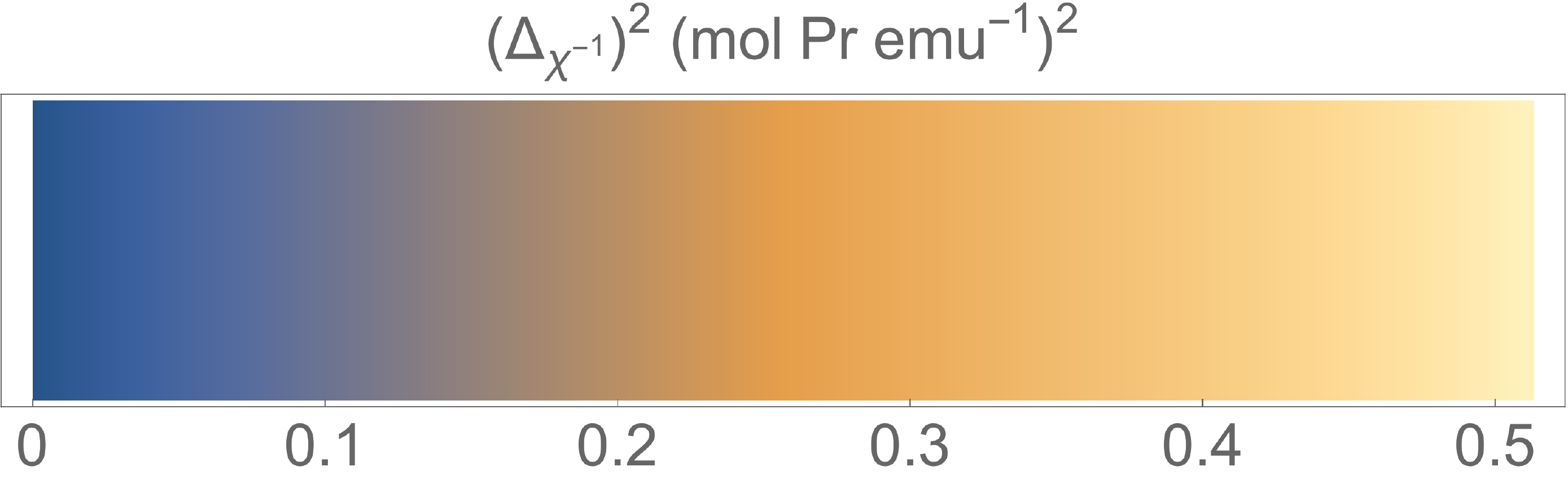}\\
\includegraphics[width=0.4\textwidth]{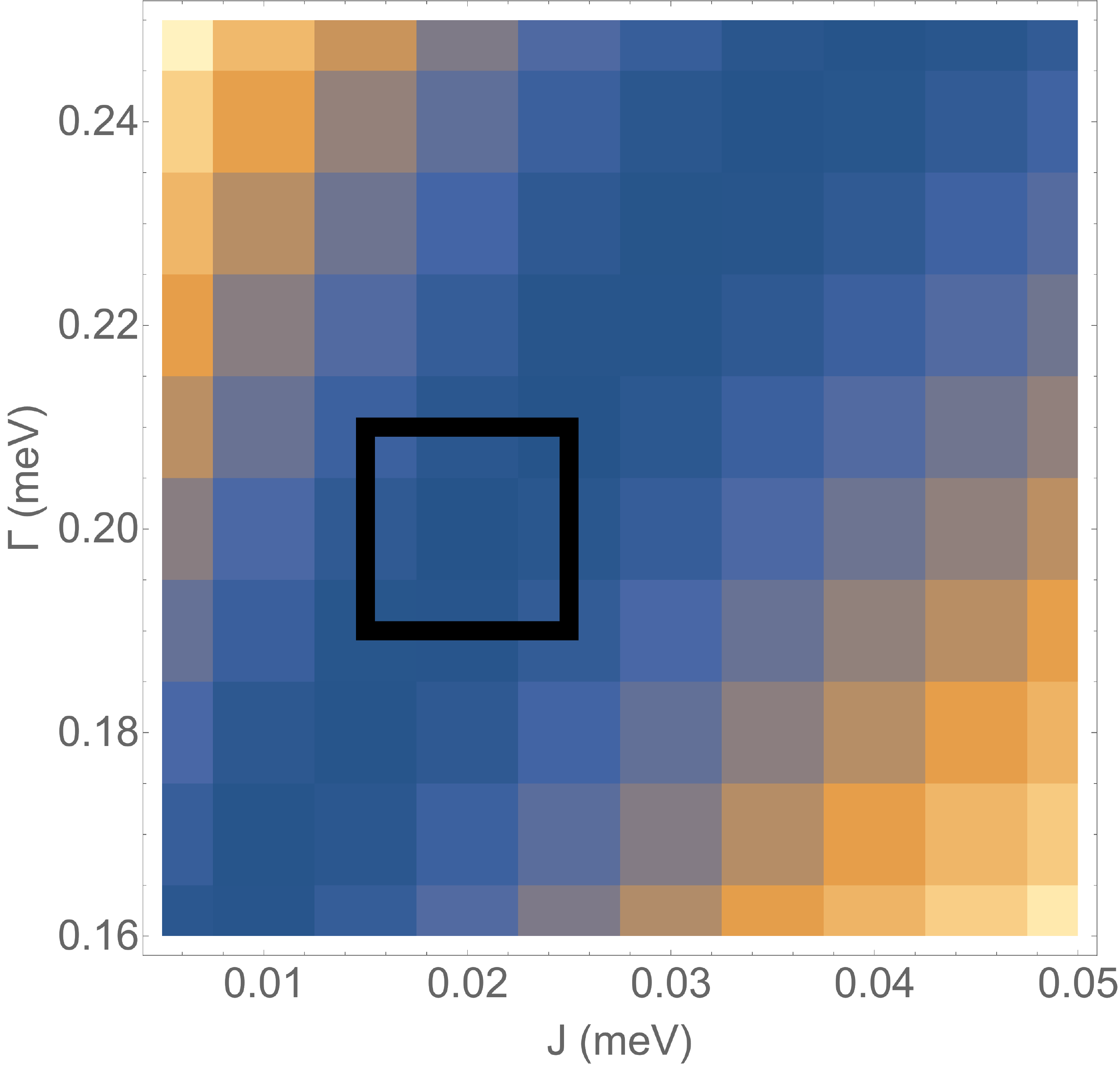}
\caption{Total squared error of the fit to the  inverse magnetic susceptbility data of Ref. 
[\onlinecite{petit16-PRB94}], as a function of the width parameter $\Gamma$
and the Ising exchange $J$. 
Here the effective magnetic moment  is fixed to $\mu_{\sf eff}=2.45 \mu_B$,
in agreement with [\onlinecite{petit16-PRB94}].
Calculations were made for each value of $\Gamma, J$ using second-order NLC
expansion and averaging over $10^6$ realizations of disorder.
There is an extended minimum in $(\Delta_{\chi^{-1}})^2$ in the $(J, \Gamma)$ plane, 
running diagonally
across the plot.
The black square indicates the region where this minimum intersects the minimum in 
$(\Delta_{C})^2$ [Fig. \ref{fig:C_quality}].
This gives us our estimate of the model parameters [Eq. (\ref{eq:estim})].
}
\label{fig:chi_quality}
\end{figure}

$\Delta_{\chi^{-1}}^2$ is plotted as a function of $J, \Gamma$ with $\mu_{\sf eff}=2.45\mu_B$
in Fig. \ref{fig:chi_quality}.
There is a line of parameter sets which each give an approximately equally good fit, diagonally
across the $(J, \Gamma)$ plane.

The black box plotted in Figs. \ref{fig:C_quality} and \ref{fig:chi_quality}
indicates the region which gives good agreement for both the heat 
capacity [Fig. \ref{fig:C_quality}] and inverse susceptibility.

\begin{figure}
\centering
\includegraphics[width=0.4\textwidth]{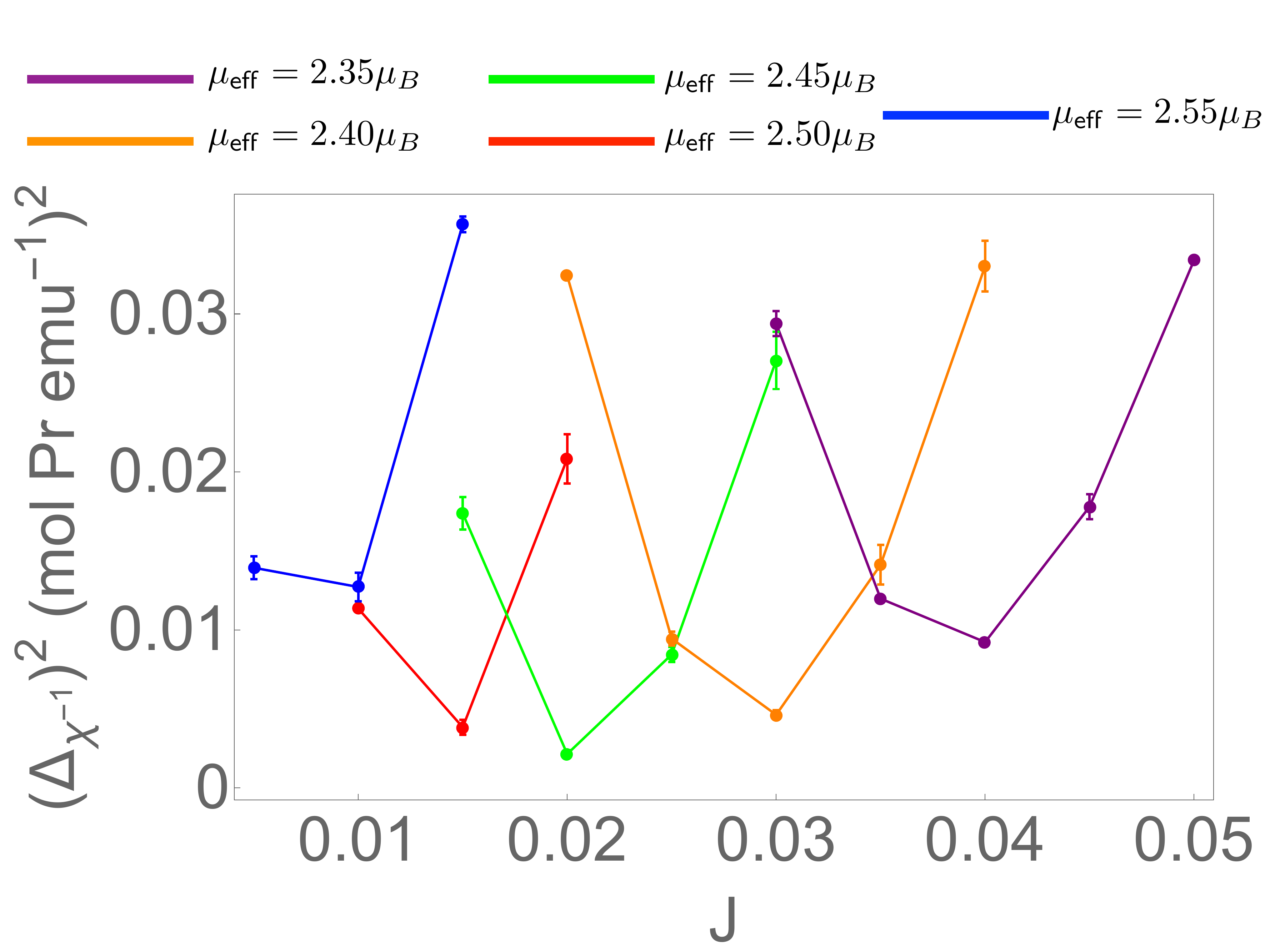}
\caption{Total squared error of the fit to the  inverse magnetic susceptbility data of Ref. 
[\onlinecite{petit16-PRB94}], as a function of the Ising exchange $J$, for
various values of $\mu_{\sf eff}$ and $\Gamma=0.20$ meV.
Calculations were made for each value of $J, \mu_{\sf eff}$ using second-order NLC
expansion and averaging over $10^6$ realizations of disorder.
Error bars are statistical errors from the disorder average.
Varying $\mu_{\sf eff}$ away from $\mu_{\sf eff}=2.45 \mu_B $ results in a 
poorer fit to the data.
 }
\label{fig:mu_quality}
\end{figure}

\begin{figure*}
\centering
\subfigure[ \ Antiferromagnetic bond $\sigma^z_k=-\sigma^z_l$]{
\includegraphics[width=\textwidth]{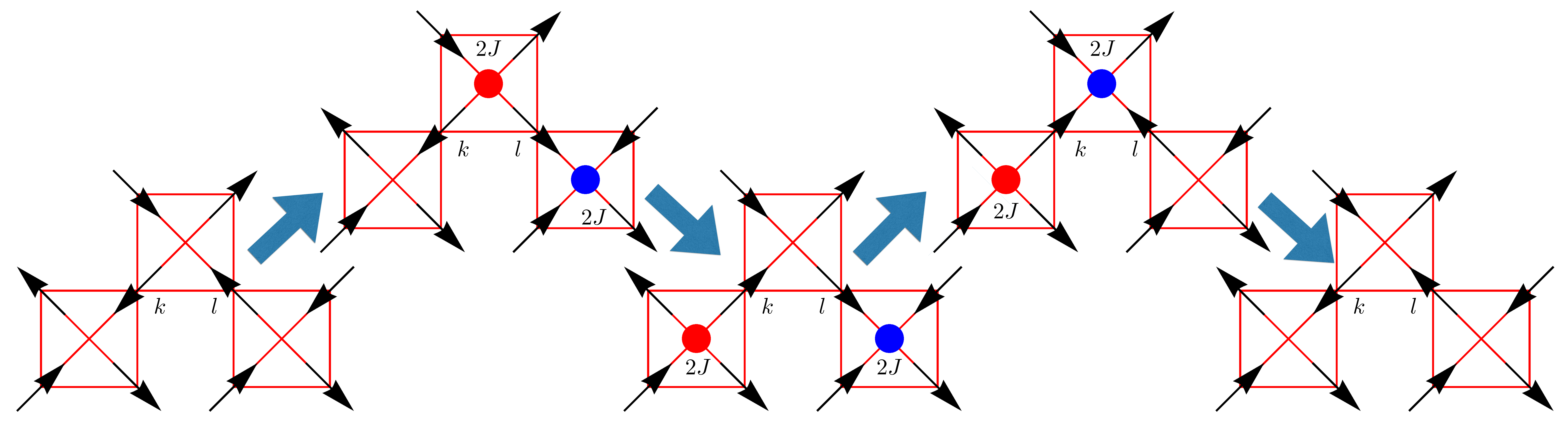}
}
\subfigure[ \ Ferromagnetic bond $\sigma^z_k=\sigma^z_l$]{
\includegraphics[width=\textwidth]{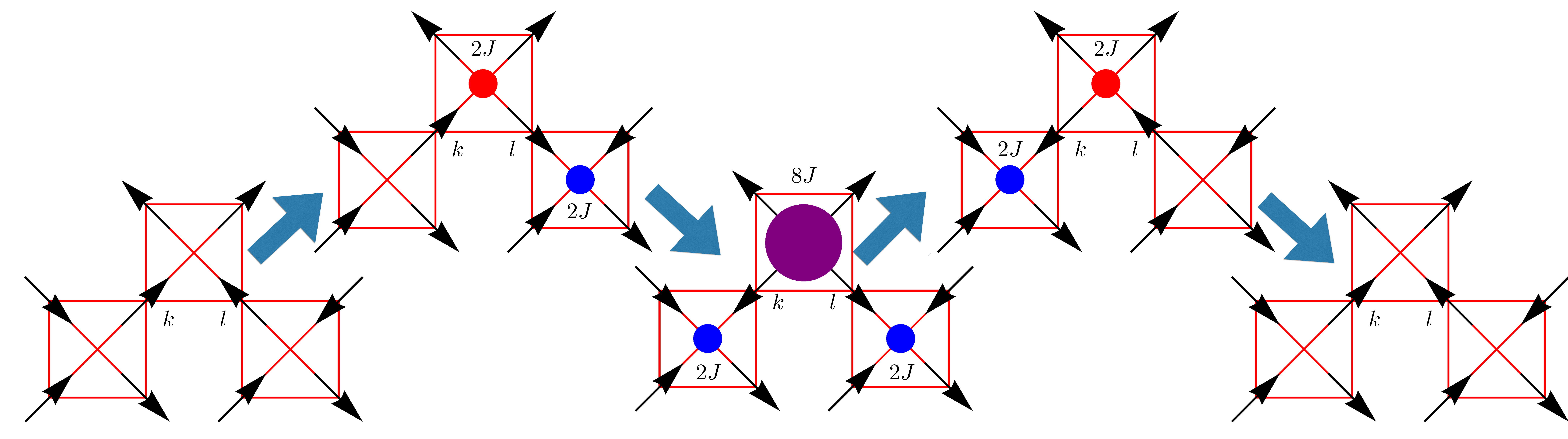}
}
\caption{
Processes contributing at fourth order in perturbation theory  [Eq. (\ref{eq:H4})]
on a bond $k, l$ in a spin ice state.
The labels on the intermediate states, indicate the energy of each
tetrahedron, relative to its ground state energy, in that state.
These quantities appear in the
denominator of the correction to the energy [Eq. (\ref{eq:H4})].
The resulting correction to the energy depends on whether the
bond $k, l$ is antiferromagnetic [(a)] or ferromagnetic [(b)]
in the initial state.
The total correction to the energy for an antiferromagnetic bond
arising from these processes is $\delta E_{\sf AFM}^{(4)}=-\frac{h_k^2 h_l^2}{16 J^3}$,
whereas for a ferromagnetic bond it is 
 $\delta E_{\sf FM}^{(4)}=-\frac{h_k^2 h_l^2}{48J^3}$.
}
\label{fig:H4processes}
\end{figure*}

This region is delineated by
\begin{eqnarray}
&&J=0.020\pm0.005 \ \text{meV}  \nonumber \\
&&\Gamma =0.20 \pm 0.01 \ \text{meV}
\label{eq:estim}
\end{eqnarray}

Lastly, we check the robustness of the parameter set against variations of 
the ordered moment $\mu_{\sf eff}$.
The heat capacity does not depend on $\mu_{\sf eff}$ so we need only check the fit 
to the susceptibility.
Fig. \ref{fig:mu_quality} shows  $\Delta_{\chi^{-1}}^2$ as a function of $J$
for values of $\mu_{\sf eff}$ in the range $[2.35, 2.55]\mu_B$, with
$\Gamma=0.2$ meV.
Moving away from $\mu_{\sf eff}=2.45\mu_B$
 significantly reduces the quality of the
optimum fit.

We estimate
\begin{eqnarray}
&&\mu_{\sf eff}=2.45 \pm 0.05 \ \mu_B.
\end{eqnarray}

\section{Confinement instability of the $U(1)$ QSL}

Here we describe the perturbation theory calculation which leads
to the result that the instability threshold for a confinement  transition of
the $U(1)$ gauge fields occurs along a line
\begin{eqnarray}
\frac{\delta h}{J} \approx \alpha \left( \frac{\bar h}{J}  \right)^3
\end{eqnarray}
with the coefficient $\alpha$ being dependent on the distribution
of transverse fields.

The confinement transition is associated with
the condensation of a dual monopole charge, and leads to
a state with frozen Ising moments $\sigma^z_i$ \cite{chen16}.
In the perturbative limit ${\bar h}, {\delta h}\ll J$, the dual
monopoles are excitations within the manifold of classical
spin ice states.
To address this instability, it is therefore appropriate to consider
perturbation theory within the manifold of classical spin ice
ground states.

Considering $\mathcal{H}_{\sf TF}$  [Eq. (4) of main text]
as a perturbation to $\mathcal{H}_{\sf SI}$  [Eq. (2) of main text]
within degenerate perturbation theory in the ice manifold, only even
orders of the expansion are non-vanishing.
At second order, there is only a trivial constant shift
in the energy 
\begin{eqnarray}
\delta E^{(2)}=-\sum_i \frac{h_i^2}{4J}
\end{eqnarray}
which is independent of the configuration of $\sigma_z^i$.

For non-uniform transverse fields, there is a non-trivial contribution arising
at fourth order.
This contribution arises from virtual processes in which two neighbouring
spins are flipped, creating excitations out of the ground state manifold,
and then both are flipped back, thus returning to the original spin configuration
[see Fig. \ref{fig:H4processes}].
Such processes generate an effective Hamiltonian, acting within the classical
ground state manifold
\begin{eqnarray}
\mathcal{H}^{g}_{(4)}=
-\mathcal{P}_{\sf g} \mathcal{H}_{\sf TF} 
\left(
\frac{\mathcal{Q}_{\sf g}}{\mathcal{H}_{\sf SI}-E_0^{\sf cl}}
\mathcal{P}_{\sf g} \mathcal{H}_{\sf TF} 
\right)^3 \mathcal{P}_g
\label{eq:H4}
\end{eqnarray}
where $\mathcal{P}_{\sf g} $ projects onto the classical ground state
manifold and  $\mathcal{Q}_{\sf g} $ projects onto its orthogonal complement.

For an antiferromagnetic bond $k, l$, with $\sigma^z_k=-\sigma^z_l$ the total contribution
to the fourth order correction to the energy is
\begin{eqnarray}
\delta E_{\sf AFM}^{(4)}=-\frac{h_k^2 h_l^2}{16 J^3}
\label{eq:DEafm}
\end{eqnarray}
while for a ferromagnetic bond we have
\begin{eqnarray}
\delta E_{\sf FM}^{(4)}=-\frac{h_k^2 h_l^2}{48 J^3}.
\label{eq:DEfm}
\end{eqnarray}

For a general bond $k, l$ we can therefore write
\begin{eqnarray}
\delta E_{k,l}^{(4)}=-\frac{h_k^2 h_l^2}{24 J^3} +\frac{h_k^2 h_l^2}{48 J^3} \sigma^z_k \sigma^z_l.
\label{eq:DEafm}
\end{eqnarray}

Including this fourth order correction, the effective Hamiltonian in the
ground state manifold now becomes an Ising model, with bond-dependent exchange
interaction
\begin{eqnarray}
&&\mathcal{H}_{\sf SI}+\mathcal{H}^{g}_{(4)}=
\sum_{\langle ij \rangle} J'_{ij} \sigma^z_i \sigma^z_j 
\label{eq:heff4}
\\
&& J'_{ij}=J+\frac{h_i^2 h_j^2}{48 J^3}
\end{eqnarray}

In the case where $h_i$ is uniform this will amount to a trivial, constant correction
to the ground state energy, as in the second order case.
However, in the disordered case where $h_i$ is non-uniform, the leading effect of
$h_i$ is to generate a correction to the effective exchange interaction such that the 
Ising exchange $J$ becomes stronger on bonds connecting pairs of sites with large
values of the transverse field $h_i$.

For a general realization of disorder, this will break the classical degeneracy of the ice
manifold and favour some frozen configuration of $\sigma^z_i$ which
minimizes Eq. (\ref{eq:heff4}). 
Such a state will prefer to have
antiferromagnetic bonds connecting sites
with large values of $h_i$.
The selection of such a frozen configuration confines the
fractional excitations of the spin liquid phase \cite{chen16}.

To estimate the instability threshold, we consider starting from the limit
$$
J\gg \bar{h}\gg \delta h
$$
and turning up the value of $\delta h$.
At  $\delta h=0$ the $U(1)$ QSL is stabilized by a ring exchange term 
with coefficient \cite{roechner16}
\begin{eqnarray}
g=-\frac{63 {\bar h}^6}{256}.
\end{eqnarray}
Increasing $\delta h$ will split the classical degeneracy of the ice manifold, via Eq. (\ref{eq:heff4}) by an
amount
\begin{eqnarray}
\epsilon \propto   {\bar h}^3 \delta h.
\end{eqnarray}
The transition from $U(1)$ QSL to frozen configuration must happen when $\epsilon \sim g$,
giving
\begin{eqnarray}
\delta h \propto h^3
\end{eqnarray}
with the coefficient of proportionality depending on the details of the transverse field distribution.

\end{document}